\documentclass[12pt]{iopart}
\def\cqfd{\hbox{}\hfill{\rlap{$\sqcup$}\rlap{$\sqcap$}}\ \ }  

\newcommand{\bn}{{\bf n}}

\newcommand{\de}{\delta}

\newcommand{\Ga}{\Gamma}

\newcommand{\Om}{\Omega}

\newcommand{\Si}{\Sigma}

\newcommand{\ra}{\rightarrow}

\newcommand{\lap}{\triangle}

\newcommand{\diag}{\mbox{diag}}

\newcommand{\be}{\begin{equation}}
\newcommand{\ee}{\end{equation}}

\newcommand{\bea}{\begin{eqnarray}}
\newcommand{\dddot}[1]{\stackrel{\cdots}{#1}}
\newcommand{\eea}{\end{eqnarray}}
\newcommand{\bean}{\begin{eqnarray*}}
\newcommand{\eean}{\end{eqnarray*}}
\newcommand{\dd}{\partial}

\newcommand{\bx}{{\mathbf x}}

\newcommand{\eg}{{\em e.g. }}

\newcommand{\VV}[1]{\overline{#1}}

\begin{document}
\title{Testing extra dimensions with the binary pulsar}

\author{Ruth Durrer$^*$ and Philippe Kocian$^\ddag$}
\address{$^*$D\'epartement de Physique Th\'eorique, Universit\'e de
  Gen\`eve, 24 quai Ernest Ansermet, CH--1211 Gen\`eve 4, Switzerland\\
$^\ddag$Institut de Physique Th\'eorique, BSP, Universit\'e de Lausanne, 
CH-1015 Lausanne,  Switzerland}
\ead{Ruth.Durrer@physics.unige.ch, Philippe.Kocian@etu.unil.ch}

\date{September 10, 2003}

\begin{abstract}
In this paper we calculate the emission of gravity waves by the binary 
pulsar in the framework of five dimensional spacetime. We consider
only  spacetimes with one compact extra-dimension. We show that the 
presence of additional degrees of freedom, especially the 'gravi-scalar' 
leads to a modification of Einstein's quadrupole formula. We compute
the induced  change for the binary pulsar PSR~1913+16 in the simple
example of a 5d Minkowski background.  In the example of a cylindrical
braneworld it amounts to
about 20\%  which  is by far excluded
by present experimental data.

\end{abstract}
\pacs{04.30.-w,04.50.+h,04.30.Db,04.80.Cc}

\maketitle

\section{Introduction}
During the last couple of years, the possibility that 
our 3+1 dimensional Universe might be a hypersurface in a higher 
dimensional spacetime has been investigated. This can be consistent
with relatively 'large' extra-dimension if only gravitational
interactions are propagating 
in the full spacetime, while gauge interactions are confined to the $3+1$ 
dimensional 'brane'~\cite{Large1,Large2}. The motivation for these
'braneworlds' is twofold. 
First of all, super string theory, which is consistently formulated only in
10 dimensions, predicts such branes as endpoints of strings, 
'D-branes'~\cite{Pol}. 
Secondly, allowing for $n$ extra dimensions of size $L$, the observed,
four dimensional Planck scale, $M_4$, relates to the fundamental Planck
scale of the underlying string theory, $M_P$  via
$M_4=M_P(LM_P)^{n/2}$. Hence for a sufficiently large length $L$ a small 
fundamental Planck scale, $M_P\sim$ TeV can lead to the observed effective 
Planck scale, $M_4\equiv 1/\sqrt{4\pi G_4} \sim 3.4\times 10^{18}$GeV,
where $G_4$ denotes Newton's constant~\cite{Large1,Large2}. A short
calculation shows that for $n\ge 2$ the required size of the extra
dimensions is less than about 1mm at which scale Newton's law has not
been extensively tested so far. Here and in 
the following we use natural units to that $\hbar=c=1$.

Motivated by this new attempt to solve the hierarchy problem, people have
started to study the modified gravitational theory obtained on the brane,
when starting from Einstein's equation in the higher dimensional 'bulk'. 
Many astrophysical and laboratory consequences of this idea have been
investigated and found to be consistent with present bounds~\cite{limits}.
For simplicity, most of the work has concentrated on 1 single extra
dimension for simplicity. We also do so in this work. 
In Refs.~\cite{bine1,bine2} the modifications of cosmological solutions are
derived. Following that, a lot of work has been devoted to cosmological 
perturbation theory in the context of brane worlds (see Refs.~\cite{Carsten}
to \cite{ours} and references therein). To do this, the authors have simply
linearized the 5-dimensional Einstein equations around a cosmological
background solution. In this paper we show that, at least for compact
braneworlds, which allow a zero--mode like it is known from
Kaluza---Klein theories, the five dimensional linearly perturbed Einstein
equations are in conflict with observations.

For this we will make use of the following very generic issue: In $d+1$ 
dimensions, the little group of the momentum state $(p,0,\cdots,0,p)$ of a 
massless particle is $SO(d-1)$ (see e.g.~\cite{Weini}). For $d=3$, the 
irreducible representations of $SO(2)$ together with parity lead to the 
usual two helicity states of massless particles, independent of their spin.
In $d=4$ dimensions however, a massless particle with spin 2 like the graviton
has the $5$ spin states coming from the five dimensional tensor representation 
 of $SO(3)$. Projected onto a $3$--dimensional hypersurface, these 
five states become the usual graviton, a massless particle with spin $1$ 
usually termed 'gravi-photon' and a massless particle with spin $0$, the 
'gravi-scalar'. These particles couple to the energy momentum 
tensor and should therefore be emitted by a time dependent mass/energy 
configuration like a binary system of heavenly bodies. This problem
has not been discussed in the literature before. In analogy to
the four dimensional quadrupole formula for graviton emission,
\be
- \dot E = {G\over 45c^5}\dddot Q_{kl} \dddot Q_{kl}, \label{quad}
\ee
one might naively expect a formula for dipole emission of for the
gravi-photon, like
\bea
- \dot E &\propto & {G\over c^3}\ddot S_l\ddot S_l  \qquad \mbox{ and}
\label{vec}\\
- \dot E &\propto & {G\over c}\dot M\dot M  \label{sca}
\eea
for the monopole emission due to the gravi-scalar respectively. Here $\bf S$
is the center of inertia,
\[ S_j = \int d^3x\; \rho(\bx)x_j \]
so that $\dot{\bf S} ={\bf P}$ and $M$ is the mass of the system. Contrary to 
$\ddot Q$, the momentum, $\dot{\bf S}={\bf P}$ and the mass are
conserved to lowest order in $G$. Therefore, 
the contributions~(\ref{vec}) and (\ref{sca}) vanish and  
the quadrupole formula, which is so well tested with the binary pulsar
PSR~1913+16~\cite{binary} that Hulse and Taylor have obtained the Nobel
price of 1993, is maintained to sufficient accuracy.

In this work we show that Eq.~(\ref{vec}) is indeed verified and there 
is no emission of gravi-photons to lowest order in $G$. However, 
Eq.~(\ref{sca}) is not. In formulating it we omitted another scalar quantity, 
the trace of the second moment on the mass distribution, $I=\int
\rho x^jx^jd^3x$.   As we shall 
derive here, this term leads to an emission of gravi-scalars 
which induce an energy loss of the same order as the usual 
quadrupole term. We will see that for the binary pulsar PSR~1913+16 this
modifies the prediction from Einstein's quadrupole formula by about 20\%
in blatant contradiction with the observations which confirm the
formula within an error of about 0.5\%.

In the next section we derive the result announced above. In Section~III we
discuss its consequences for models with one extra--dimension.

Throughout  we use capital Latin letters for five dimensional 
indices $0,1,2,3,4$, Greek letters for four dimensional ones and lower 
case Latin letters for three dimensional spatial indices. The spatial 
Laplacian is denoted by, $\de^{ij}\dd_i\dd_j \equiv \lap$, and spatial
 vectors are indicated in bold face.

\vspace{5mm}

\section{The modified quadrupole formula for spacetimes with one
  extra--dimension} 

In order to be specific and for simplicity, we consider a $4+1$
dimensional cylinder, with a rolled up fourth spatial dimension of
length $L$. Since this length which is of 
the order of micrometers or smaller, is much smaller that the scale of the
$3+1$ dimensional system which is emitting gravity waves, we shall
only consider states which 
are zero-modes with respect to the fourth spatial dimension.

We choose a gauge such that the perturbed five dimensional line element is 
given by~\cite{ours}
\bea
ds^2 &=& -(1+2\Psi)dt^2 -2\Si_idtdx^i  -2{\cal B}dtdy 	+2{\cal E}_idx^idy 
\nonumber \\  &&   +\left[(1-2\Phi)\de_{ij}+2H_{ij}\right]dx^idx^j 
		 +(1+2{\cal C})dy^2  \nonumber \\
 &=& (\eta_{AB}+h_{AB})dx^Adx^B~. \label{metric}
\eea
Here $\Si_i$ and ${\cal E}_i$ are divergence free vectors, $H_{ij}$ is a 
traceless, divergence free tensor and $\eta_{AB}$ is the five
dimensional Minkowski metric. In the appendix we show, that the
variables defined above are gauge-invariant combinations of the most
general metric perturbations. There we also see that the gauge choice
made in Eq.~(\ref{metric}) fixes the gauge completely. The
extra-dimension is parameterized  
by the co-ordinate $y$, $0\le y\le L$ for our cylindric spacetime. We
are working within linear perturbation theory and therefore require
$|h_{AB}| \ll 1$. The 
source free linearized Einstein equations for this geometry can be reduced to
three wave equations,
\bea
(\dd_t^2 -\dd_y^2-\lap)\Phi &= &0 ~, \label{sourcefree1} \\
(\dd_t^2 -\dd_y^2-\lap)\Si_i &= &0 ~,\\
(\dd_t^2 -\dd_y^2-\lap)H_{ij} &= &0 ~.
\eea
The other variables are then determined by constraint equations, 
\bea
\lap {\cal C} &=& (2\lap + 3 \dd_y^2)\Phi \\
\Psi &=&  \Phi - {\cal C} \\
\lap {\cal B} &=& -6\dd_t\dd_y\Phi\\
\dd_y{\cal E}_i &=& -\dd_t\Si_i \label{sourcefreelast}
\eea
(for more details see the Appendix and~\cite{dip}). The important
point is that the above wave equations describe simply
the five propagating modes which we also expect from our group theoretical 
argument above. We now consider a four dimensional, Newtonian source 
given by
\bea
T_{\mu 4} &=& T_{44} = 0  \label{T4}\\
T_{00} &=& \rho(\bx,t)\de(y)\\
T_{0i} &=& \big(\dd_iv+ v_i\big)(\bx,t)\de(y)\\
T_{ij} &=& \left[P\de_{ij} + (\dd_i\dd_j-{1\over 3}\de_{ij}\lap )\Pi 
        ~  +  \right.\nonumber \\
&& \left. +  {1\over 2}\left(\dd_i\Pi_j+\dd_j\Pi_i\right)
	  +\Pi_{ij}\right](\bx,t)\de(y)  \label{Pi}
\eea
where $v_i$ and $\Pi_i$ are divergence free vector fields and $\Pi_{ij}$ 
is a traceless divergence free tensor in three spatial dimensions.
Eqs.~(\ref{T4}) to (\ref{Pi}) represent the most general energy
momentum tensor on the brane with vanishing energy flux into the
bulk. However, being interested in the binary pulsar, we shall
restrict our attention to Newtonian sources, i.e. sources with
$\rho= |T_{00}|\gg |T_{0i}|\sim \rho V $  , $|T_{00}|\gg  |T_{ij}\sim
\rho V^2$, where $V \ll 1$ is a typical velocity of the system.

Since we are only interested in the zero-mode with respect to the
fourth spatial dimension, we may integrate over the fifth
dimension. The four dimensional linearized Einstein equations with
source term then become  (see Appendix)
\begin{eqnarray}
\label{eq:scal1'}
\mbox{scalar perturbations:}\nonumber\\
\lap (2 \Phi -{\cal C}) & = & 8 \pi G_4 \rho \\
\label{eq:scal2'}
\partial_t {\cal C} - 2 \partial_t \Phi & = & 8 \pi G_4 v \\
\label{eq:scal3'}
{\partial_t}^2 (-{\cal C} + 2 \Phi) & = & 8 \pi G_4 \Big(P + \frac{2}{3} \lap \Pi \Big) \\
\label{eq:scal4'}
\Phi - \Psi -{\cal C} & = & 8 \pi G_4 \Pi \\
\label{eq:scal5'}
\lap {\cal B} & = & 0~, \qquad
\partial_t {\cal B}  =  0 \\
\label{eq:scal7'}
- \lap (2 \Phi - \Psi) + 3 {\partial_t}^2 \Phi & = & 0 \\
\mbox{vector perturbations:}\nonumber \\
\label{eq:vec1'}
 \frac{1}{2} \lap {\Sigma}_i & = & 8 \pi G_4 v_i \\
\label{eq:vec2'}
\partial_t {\Sigma}_i & = & 8 \pi G_4 \Pi_i \\
\label{eq:vec3'}
 \lap {\cal E}_i - {\partial_t}^2 {\cal E}_i & = & 0 \\
\mbox{tensor perturbations:}\nonumber \\
\label{eq:tens'}
- \lap H_{ij} + {\partial_t}^2 H_{ij} & =
  & 8 \pi G_4 \Pi_{ij} \ . 
\end{eqnarray}
where $G_4=(\pi/2)L^{-1}G_5$ is the four dimensional Newton constant,
$M_4=(2\pi^2 G_5/L)^{-1/2}$. Here we have used the relation $G^{-1}_d =
M_d^{d-2}{\rm vol}({\bf S}^{d-2}) =
M_d^{d-2}2\pi^{d-1\over2}/\Ga({d-1\over 2})$ between the
$d$-dimensional Newton constant and Planck mass. This relation ensures
that Gauss' law is true in any dimension. Einstein's equations in an
arbitrary number of dimensions are $G_{AB}=2M_d^{2-d}T_{AB}$.
Note that the equations for the zero-mode are very similar to the
Kaluza-Klein approach. Both $\cal B$ and ${\cal E}_i$ are completely
decoupled. In Kaluza-Klein theories they play the role of the
electromagnetic field which couples to the current $J_\mu \propto
T_{\mu 4}$ which we have set to zero in this work. In a coordinate
system where the brane is fixed 
(no brane bending) the only difference for the zero-mode, in our
'test--brane' analysis is the
relation between the four-- and five--dimensional gravitational constant.
From  Eqs.~({\ref{eq:scal1'}) to (\ref{eq:tens'}) we can derive three
  wave equations with source term: 
\bea
(\dd^2_t-\lap)\Phi &=& -8\pi G_4(\rho -\lap\Pi) \label{sca2}\\
(\dd^2_t-\lap)\Si_i &=& 8\pi G_4(\dot\Pi_i +2v_i) \label{vec2}\\
\dd_t\Si_i &=& 8\pi G_4\Pi_i \label{vec3}\\
(\dd^2_t-\lap)H_{ij} &=& 8\pi G_4\Pi_{ij}~, \label{ten2}
\eea
The other variables are again determined by constraint equations.

Eq.~(\ref{ten2}) leads over well-known steps to the quadrupole formula for 
Newtonian sources. The retarded solutions of the wave 
Eqs.~(\ref{sca2}) and (\ref{vec2}) determine $\Si_i$ and $\Phi$ far away from 
the source. It is easy to see that the a priori dominant terms $\rho$ in 
(\ref{sca2}) and $v_i$ in (\ref{vec2}) lead exactly to the terms predicted 
in Eqs.~(\ref{sca}) and (\ref{vec}). Hence their contribution to the
energy emission vanishes (to lowest order) by mass and momentum 
conservation. We therefore concentrate on the terms coming from the 
stress tensor. Energy momentum conservation ($T^{\mu\nu},_\mu=0$) gives
\be \lap(\lap\Pi +{3\over 2}P) = {3\over 2}\dd_i\dd_jT^{ij} =
{3\over 2}\dd_t^2\rho~,\ee
so that
$ \lap\Pi= {3\over 2}(\ddot U - P)$, where 
\be U(t,\bx) = {1\over 4\pi}\int_{\Re^3}{\rho(t,\bx') \over |\bx-\bx'|}d^3x' ~.\ee
Taking the divergence of (\ref{Pi}) and using the above expression for
$\lap\Pi$  we find
\be \lap\Pi_j = 2\dd_i(T^i_j -\de^i_j\ddot U) ~,\ee
so that, to order $1/r$, each component of the vector potential
$\Pi_i$ is the integral 
over a divergence which vanishes for a confined source. Hence
$\Pi_i\propto r^{-2}$; together with (\ref{vec3}) this implies
that also $\Si_i$ is of order $r^{-2}$. The  energy flux from
$\Si_i$ is expected to be proportional to  $t_{0j} \propto (\dd_t\Si_i)
(\dd_{[i}\Si_{j]}) \propto 1/r^4$ (the bracket $[..]$ indicates
anti-symmetrization) which shows that we have no energy emission
from the  gravi-photon far away from the source. 

The situation is different for the gravi-scalar, $\Phi$.
There the energy flux is proportional to 
$t_{0i}\propto \dd_t\Phi\dd_i\Phi$. To order $1/r=1/|\bx|$, the scalar field
$\Phi$ is given by
\be 
\Phi(t,\bx) = {G_4\over r}\int_{\Re^3}\!\left[3(\ddot U(t',\bx') - P(t',\bx'))-
2\rho(t',\bx')\right]d^3x'. \label{Phi}
\ee 
Here $t'=t-|\bx-\bx'|$ is retarded time.   
The last term does not contribute to $t_{0i}$ since $\dd_t\int\rho =\dot M =0$.
Furthermore, $\dot \rho =\dd_iT^{0i}$ so that
\bea
4\pi\dot U(t,\bx) &=& \int_{\Re^3}{\dd'_i T^{0i}(t,\bx') \over
  |\bx-\bx'|}d^3x'
= - \int_{\Re^3} T^{0i}(t,\bx')\dd'_i{1 \over
  |\bx-\bx'|}d^3x' \\
&=&-\dd_i\int_{\Re^3}{T^{0i}(t,\bx') \over
  |\bx-\bx'|}d^3x' ~.
\eea
Here $\dd'_i$ denotes differentiation w.r.t. $\bx'$ while  $\dd_i$
is differentiation w.r.t. $\bx$.   
In other words, $\dot U$ is a divergence and therefore does not contribute
in the integral (\ref{Phi}). 

Finally, we use the identity (see~\cite{strau}) 
\[ 3 \int_{\Re^3}P =  \int_{\Re^3} \de^{ij}T_{ij} = 
 {1\over 2} \int_{\Re^3} \de^{ij}\ddot\rho x'_ix'_j \]
which is a simple consequence of energy momentum conservation and Gauss' law. 
All this leads to
\bea
 \dot\Phi(t,\bx) &=& {G_4\over 2r}\int\dddot\rho(t',\bx')x'^2d^3x' +{\cal
   O}\left({1 \over r^2}\right) \label{Phidot}\\
\mbox{and }\quad &&\nonumber\\
\dd_i\Phi(t,\bx) &=& {G_4n_i\over 2r}\int\dddot\rho(t',\bx')x'^2d^3x' +{\cal
   O}\left({1 \over r^2}\right). \label{Phii}
\eea
Here $n^i=n_i=\bx^i/r$.
The energy emission from the gravi-scalar is then given by
\bea
   -\left.\left({dE\over dt}\right)\right|_{\rm scalar} 
   &=&\lim_{r\ra\infty}\int_{{\cal S}^2}\! r^2 d\Om n^it_{0i}(r\bn)
   \nonumber\\
&\propto & \lim_{r\ra\infty}\int_{{\cal S}^2}\! r^2 d\Om
n^i\dot\Phi\dd_i\Phi = \pi G_4^2\left[{d^3\over dt^3} \int \rho x^2d^3x
    \right]^2~.   
\label{en1}
\eea
For the last equality sign we have used the results (\ref{Phidot}) and
(\ref{Phii}) and integrated over the sphere.

To obtain  the proportionality factor which is still missing in
Eq.(\ref{en1}), we write the five 
dimensional action up to second order in the scalar metric perturbations
\bea
 S &= & {1\over 8\pi^2 G_5}\int_0^L\! dy \int_{\Re^4}\! d^4x\sqrt{|g|}R 
\nonumber\\
 &=&  {1\over 16\pi G_4}\int_{\Re^4}\! d^4x(1+{1\over 2}h^A_A)(R^{(1)} +R^{(2)})
\nonumber\\
&=& \int_{\Re^4}\! d^4x{1\over 2}\dd_\mu\varphi\dd^\mu\varphi   + \mbox{
  a total derivative} ~.  \label{vphi}
\eea
Here $h^A_A=2(\Psi - 3\Phi + {\cal C})$ is the trace of the  metric
perturbations and $R^{(1)}$ and $R^{(2)}$ are the first and second
order (scalar) perturbations of the five dimensional Riemann scalar.
The last equation is just the ansatz for a
usual, canonically normalized scalar field action. The source free
Einstein equations for the zero-mode can be obtained by varying this
action.  Therefore, the energy flux of the mode is given by 
$t_{0i}=\dd_0\varphi\dd_i\varphi$. A straight forward
but somewhat tedious calculation of $R^{(1)}$ and $R^{(2)}$, in which
we use the homogeneous constraint equations to express all scalar
perturbations variables in terms of $\Phi$, gives finally
\[ \varphi = \sqrt{21\over 8\pi G_4}\Phi \quad \mbox{ so that }\quad 
 t_{0i} = {21\over 8\pi G_4}\dot\Phi\dd_i\Phi~.\]
Inserting this proportionality factor in Eq.~(\ref{en1}) leads to
\be
-\left.\left({dE\over dt}\right)\right|_{\rm scalar} = {21G_4\over 8}
 \left[{d^3\over dt^3}
 \int_{\Re^3}\rho x^2\right]^2 ~. \label{enfin}
\ee

A textbook calculation for a binary system in Keplerian orbit, like it 
is presented \eg in Ref.~\cite{strau},  now gives for
the energy loss averaged over one period of the system
\bea
- \left\langle{dE\over dt}\right\rangle &=& 
  {21 G_4^4M_1^2M_2^2(M_1+M_2)g(e)\over 4 a^5(1-e^2)^{7/2}}  \label{scalquad}\\
\mbox{with } ~ \quad g(e) &=& e^2(1+e^2)~.
\eea
Here $M_{1,2}$ are the two masses of the system, $a$ is the semi-mayor axis 
and $e$ is the eccentricity of the orbit. Comparing this with the
quadrupole formula for a Keplerian binary system which can be found
e.g. in~\cite{strau}), we see that (\ref{scalquad}) agrees with it
upon replacing  
${21g(e)\over 4}$ by $32f(e)\over 5$ with $f=1 +(73/24) e^2 +(37/96)e^4$.
Inserting the value $e=0.617$ for PSR~1913+16 given in~\cite{binary},
one finds that this new contribution amounts to 19.9\% of the ordinary
quadrupole prediction. 
Therefore, the slowdown of the period of PSR~1913+16 should be
\bea
 \dot T|_{\rm tens+scal} &=& -2.88\times 10^{-12} \quad \mbox{ instead of }\\
\dot T|_{\rm tens} &=& -2.4024\times 10^{-12} \label{tenb}  ~.
\eea
But the experimental value~\cite{binary} is in perfect agreement 
with the quadrupole formula~(\ref{tenb}) and contradicts such a huge
correction: 
\be \dot T|_{\rm exp} = -(2.408\pm 0.01)\times 10^{-12} ~.\ee

Our result is not entirely self-consistent in that we assume that the
masses $M_1$ and $M_2$ inferred for the two neutron stars are the same
as in ordinary four dimensional gravity. This is not necessarily the
case and probably also depends on the method by which they are
determined. In principle we would have to specify two additional
measurements and check whether different masses $M_1'$ and $M_2'$ can
be found so that all the experiments agree with the results from four
dimensional gravity for the masses $M_1$ and $M_2$. This procedure has
been undertaken in the work on general scalar-tensor theories of
gravity in Ref.~\cite{DaEs}. Clearly, it would be a very surprising coincidence
if such masses could be found. 

Finally, we want to mention that other tests, like for example light
deflection around the sun most probably also lead to deviations from
the four dimensional result. 
\vspace{8mm}

\section{Conclusion}

Our calculations show clearly that a $4+1$ dimensional cylindrical spacetime
which satisfies the five dimensional Einstein equations contradicts
observations. This finding is independent of the size of the extra dimension. 
Note that we did only take into account the Kaluza-Klein zero-mode and 
hence the size $L$ of the extra dimension just enters in the relation 
between the five and four dimensional gravitational constants, $G_4
=G_5\pi/2L$.  It is important to note at this point that the zero-mode
in general of course depends on the background geometry and thus on
the warp factor (it is a solution to the five dimensional
d'Alambertian, ${\cqfd}_5$ with respect to the background geometry). In
our case, the background geometry is trivial. If the  warp factor
depends on the size of the extra-dimension this will of
course also imprint itself on the zero-mode. 

In our calculation we have considered a simple cylindrical
spacetime. But our result would not alter substantially if we would
consider a more realistic, warped spacetime which corresponds to an expanding
universe. The mayor exception are warped models with {\em non-compact}
extra-dimensions like the Randall Sundrum~II model~\cite{RS2}, where
the scalar gravity mode  is not normalizable and therefore our
discussion does not apply.
But in all models with {\em compact} extra-dimensions, for example in all
models with two branes, like Randall 
Sundrum~I~\cite{RS1} or the ekpyrotic universe~\cite{ekp} the
mode discussed here is normalizable. Its coupling is determined by the  five
dimensional Einstein equations. We therefore expect that (besides all
the other problems of scalar--tensor theories of gravity~\cite{DaEs})
this mode be excited  by a system like the binary pulsar and 
lead to deviations from Einstein's quadrupole formula. 
Even though the exact amount of change will depend on the details of
the model, generically we can expect it to be of the same order of magnitude.
 The expansion of the Universe, which determines the jump of the
extrinsic curvature between the two sides of the brane is not relevant
on scales of a binary pulsar and so we expect a realistic braneworld
model to lead to very similar results. The possibly more complicated
geometry of the extra-dimension may somewhat affect the numerical
result but as long as  the {\em zero-mode} discussed here exists
({\em i.e.} as long as it is normalizable) the problem will remain.

In order to find consistent four dimensional Einstein gravity on the brane,
the  five dimensional Einstein equations have to be modified somehow 
 at very low energy. One possibility might be that the gravi-scalar 
$\varphi$ acquires a small mass  which is larger than the frequency of 
the binary pulsar,
$m_\varphi > 1/T \simeq 3.6\times 10^{-5}$Hz $\simeq 2.4\times 10^{-20}$eV
(e.g. via the Goldberger--Wise mechanism~\cite{GW}). Such a tiny mass
would already  suffice to inhibit the field 
$\varphi$ to be 
emitted by the binary pulsar PSR~1916+13. To interpret the gravi-scalar 
physically, it is useful to notice that in source free space, for the zero modes
we have $2\Phi = C$. But $C$ determines the size of the extra dimension,
$L+\de L = \int_0^L(1+C)dy$.  Therefore, if one wants to stabilize the size 
of the extra dimensions one needs a positively curved potential, hence a mass,
for $C$ and thus for $\Phi$. This argument is well known in string
inspired Ka\-lu\-za-Klein theories. There, the size of the extra
dimensions is given by the string scale, $\ell$ and T-duality, the duality of
string theory under the transformation 
$r\ra \ell^2/r$, indicates that a potential for $\de
L$ and hence for $C$ should have a minimum at $L=\ell$, $C=0$, in
other words a positive mass term, see \eg~\cite{gabi}. 
Some non-gravitational (stringy) correction must provide this mass.

But even if we
set $C=0$, the scalar perturbation equations~(\ref{eq:scal1'}) to 
(\ref{eq:scal7'}) do not reduce
to the usual four dimensional perturbation equations: using
(\ref{eq:scal3'}) and the Laplacian of (\ref{eq:scal4'}) to eliminate
$\dd_t^2\Phi$ and $\lap(\Phi-\Psi)$ in (\ref{eq:scal7'}), we obtain
\be \lap\Phi=8\pi G_4 {3\over 2}P \ee
Comparing this with Eq.~(\ref{eq:scal1'}) we find that these equations
are only compatible when $\rho=3P$, in other words, if the matter is
relativistic. This shows that the problem discussed in this paper goes
beyond the radion problem: even if the radion $\cal C$ is fixed, we do
not recover the four dimensional Einstein equations. 

Moreover, if non-gravitational corrections influence the evolution
equation of the gravi-scalar, they
may very well also introduce modifications to the interactions
on the brane which are usually inferred  from the five dimensional Einstein
equations, like the Israel junction conditions~\cite{Israel}, or
certainly the gravitational perturbation equations as derived in
Refs.~\cite{Carsten} to \cite{ours}. 

As mentioned above, the situation is mitigated if the geometry is warped, like
in the Randall-Sundrum model~\cite{RS2}. In such models, the fifth
dimension may be non-compact and the gravi-scalar zero-mode can be 
non-normalizable in the sense that the integral $\int\sqrt{|g|}\varphi^2 dy
= \infty$. Then the gravi-scalar cannot be excited as it has infinite
energy. 
Within the language of classical relativity, we may simply say that
the 'gravi-scalar mode' turns out to be large for some values of $y$
and  linear perturbation theory is therefore not applicable.
 
Let us finally summarize our main conclusion: If a model with extra dimensions
is such that the gravi-scalar coming from Einstein's equations in five
dimensions is normalizable, which is always the case if the
extra-dimensions are compact, non-gravitational interactions need to
intervene in order to make the model compatible with observations. 
They have to give the gravi-scalar a mass.
In such models therefore, we cannot simply use the five dimensional
Einstein equations to infer the four dimensional equations of motion. 
Especially, we cannot apply five dimensional gravitational
perturbation theory to, e.g., a cosmological solution in order to learn
about four dimensional cosmological structure formation.
The perturbation theory derived in Refs.~\cite{Carsten} to \cite{ours} can at
best be applied to models where the gravi-scalar is not normalizable.

Clearly, the situation does not improve if there are more than one
extra dimensions. In this case there are several gravi-scalars which
may all contribute a similar amount to the emission of gravity waves.

\ack
We acknowledge discussions with Gilles Esposito-Farese, Misha
Shaposhnikov and Gabriele Veneziano.  
This work is supported by the Swiss National Science Foundation.

\appendix

\section{ \\ Gauge--invariant perturbation equations for a 5d Minkowski
  braneworld} 

We consider a five dimensional Minkowski bulk with background metric
$(\eta_{AB}) =\diag(-1,1,1,1,1,)$. The most general perturbed metric
is then of the form
\bea
d s^2 &=& -(1 + 2 A ) dt^2 - 2  S_i dt dx^i
   +  (\de_{ij} + h_{ij}) dx^i dx^j
   - 2  B d y  dt  \nonumber \\   &&
+ 2 E_{ i} dy dx^i + (1+ 2 C) dy^2.
\label{mpert1}
\eea

The spatial vectors $S_i$ and $E_i$ can be decomposed into scalar
(spin zero) and vector (spin one ) components and the  spatial
tensor $h_{ij}$ can be decomposed into scalar, vector and tensor (spin
two ) components;
\begin{eqnarray}
S_i  =  \nabla_i S + \VV{S}_i , & \quad & E_i  =  \nabla_i E +
\VV{E}_i ,\\
h_{ij}  =   2 H \gamma_{ij} + 2 F_{ij} , & \quad & 
F_{ij}  =  \nabla_{(i} F_{j)} + H_{ij} ,  \quad  
F_i  =  \nabla_i F + \VV{F}_i . \label{mpertfin}
\end{eqnarray} 
Here $\VV{E}_i$, $ \VV{F}_i$ and  $ \VV{S}_i$ are divergence free vectors
and $H_{ij}$ is a divergence free, traceless symmetric tensor.
To first order, scalar vector and tensor perturbations evolve
independently. Spatial indices can be raised and lowered with the
background metric $\de_{ij}$.

Let us now study the transformation of the defined perturbation
variables under infinitesimal coordinate transformations (gauge
transformations) in the bulk.
We consider an infinitesimal coordinate transformation
\be
x^A \to x^A + \xi^A,  \label{infi}
\ee
where we set
\begin{equation}
(\xi^A)  =  (T, L^i, L^4) ,\quad L^i  =  \nabla^i L + \VV{L}^i .
 \label{cico}
\end{equation}
The three scalar ($T,~L,~L^4$) and one vector type ($\VV{L}^i$)
gauge-transformation allow us to 
gauge to zero three scalar and one vector variable. Instead of
choosing a specific gauge (like in the main text of this paper) we
shall define gauge invariant combinations and express the first order
Einstein equations in terms of these. This is always possible as a
consequence of the Steward-Walker Lemma~\cite{StWa}.

Under the above coordinate change, the geometrical  perturbations
transform according to the Lie derivative of the background metric,
$g_{AB} \ra g_{AB} +L_\xi\eta_{AB}$ 
\begin{eqnarray}
A  \to  A + \dd_t T, & \quad & 
\VV{S}_i  \to  \VV{S}_i - \dd_t \VV{L}_i  , \label{transA}\\
H  \to  H , & \quad &
\VV{F}_i  \to  \VV{F}_i + \VV{L}_i ,  \label{ApFi}\\
H_{ij} \to  H_{ij} , & \quad &
B  \to  B - \dd_t L_4 + \dd_y T , \\
\VV{E}_i  \to  \VV{E}_i + \dd_y\VV{L}_i , & \quad &
C \to   C + \dd_yL_4 , \label{transepp}\\
S   \to  S   + T -\dd_t L,  & \quad &
F  \to  F  + L , \\
E \to E + \dd_yL + L_4  . && \label{ApF}
\end{eqnarray}
We can therefore define the following four scalar and two vector
perturbation variables which are gauge
invariant,
\begin{eqnarray}
\Psi & = & A  -\dd_t(S +\dd_t F)  , \label{vPsi} \\
\Phi & = & -H , \label{vPhi}\\
{\cal B} & = & B - \dd_y(S +\dd_t F) + \dd_t(E -\dd_y F) , \label{vvp} \\
{\cal C} & = & C - \dd_y (E - \dd_y F) , \label{vhpp}\\
\Si_i & = & S_i + \dd_t\VV{F}_i , \label{vvi}\\
{\cal E}_i & = & \VV{E}_i - \dd_y \VV{F}_i \label{vhci}.
\end{eqnarray}
The tensor variable $H_{ij}$ is gauge invariant since there are no
tensor type gauge transformations.

The stress energy tensor defined in Eqs.~(\ref{T4}) to (\ref{Pi})
is a first oder perturbation with vanishing background component and
it is therefore gauge invariant by itself. This is a specialty of the
Minkowski background. The much more complicated general
gauge-invariant  variables for a generic evolving bulk can be
found in Ref.~\cite{ours}.

From Eqs.~(\ref{ApFi}) and (\ref{ApF}) it is clear that the
gauge is completely fixed by setting $F=S=E=0$ and $\VV{F}_i=0$. This
'generalized longitudinal gauge'  is
precisely the gauge choice adopted in the main text of this paper.

We can now go on and compute the components of the curvature in terms
of our perturbation variables. In Ref.~\cite{ours} this is done for
the generic evolving case. Here we repeat for convenience the
expressions for the Christoffel symbols and for the Einstein tensor in
our simple case. In contrast to the generic situation, the Einstein
tensor is gauge invariant in for a Minkowski bulk.

\subsection*{Christoffel symbols}

\begin{eqnarray}
 \Gamma^{\;0}_{0 0}  =  \dd_t A  , & ~ &
 \Gamma^{\;0}_{0 i}  =   \dd_i A  , \\
 \Gamma^{\;0}_{i j}  = 
  \nabla_{(i} S_{j)} + {1\over 2} \dd_t h_{i j} , & ~ &
 \Gamma^{\;0}_{0 4}  =  \dd_y A  , \\
 \Gamma^{\;0}_{i 4}  =  {1\over 2}\left(
   \dd_y  S_i + \dd_i B + \dd_t E_i \right) , & ~ &
 \Gamma^{\;0}_{4 4}  =  \dd_y B + \dd_tC ,\\
 \Gamma^{\;i}_{0 0}  =  \dd^i A - \dd_t S^i , & ~ &
 \Gamma^{\;i}_{0 j}  =  {1\over 2}\left(
  \dd^i S_j - \dd_j S^i + \dd_t h^i_j \right) , \\
 \Gamma^{\;i}_{j k}  =  {1\over 2} 
 (\dd_j h^i_k + \dd_k h^i_j - \dd^i h_{jk}) , & ~ &
 \Gamma^{\;i}_{0 4}  =  {1\over 2} \left( \dd^i B 
 \! - \! \dd_y S^i \! + \! \dd_t E^i \right) , \\
 \Gamma^{\;i}_{j 4}  =  {1\over 2}\left(
   \dd_y h^i_j - \dd^iE_j + \dd_j E^i \right)  , & ~ &
 \Gamma^{\;i}_{4 4}  =   \dd_y E^i - \dd^i C ,\\
 \Gamma^{\;4}_{0 0}  =  \dd_y A  - \dd_t B , & ~ &
 \Gamma^{\;4}_{0 i}  =  {1\over 2} \left(
   \dd_y S_i  -  \dd_i B + \dd_t E_i \right) , \\
 \Gamma^{\;4}_{i j}  =  -{1\over 2} \dd_y h_{i j} 
                + \dd_{(i} E_{j)} , & ~ &
 \Gamma^{\;4}_{0 4}  =   \dd_t C , \\
 \Gamma^{\;4}_{i 4}  =  \dd_i C  , & ~ &
 \Gamma^{\;4}_{4 4}  =   \dd_y C ,
\end{eqnarray}

Here $X_{(i}Y_{j)} ={1\over 2}(X_iY_j +X_jY_i)$ indicates
symmetrization in the indices $i$   and $j$.

\subsection*{Einstein tensor}

Over the Ricci tensor one can now compute also the Einstein
tensor. This is a somewhat lengthy but straight forward
calculation. Also here, the more complicated results for an evolving
bulk can be found in Ref.~\cite{ours}

\begin{eqnarray}
G_{00}  & = &  \lap (-{\cal C}+2\Phi) +3\dd_y^2\Phi  , \\
G_{0i} & = & \dd_i\left[\dd_t(2\Phi -{\cal C}) -{1\over 2}\dd_y{\cal B}\right]
 +{1\over 2}\dd_y\left[\dd_y\Si_i + \dd_t{\cal E}_i\right] +{1\over 2}\lap\Si_i
 ,\\ 
G_{ij} & = &(\lap\de_{ij}-\dd_i\dd_j)\left(\Psi -\Phi +{\cal C}\right)
+\de_{ij}\big[\dd_t^2(-{\cal C}+ 2\Phi) +\dd_y^2(\Psi-2\Phi)
 \nonumber\\ 
 & &  -\dd_t\dd_y{\cal B} \big]  + \dd_t\dd_{(i}\Si_{j)} 
    + \dd_y\dd_{(i}{\cal E}_{j)} +(\dd_t^2-\lap-\dd_y^2)H_{ij} ,\\ 
G_{04} & = &{1\over 2}\lap{\cal B} + 3\dd_t\dd_y\Phi , \\
G_{i4} & = & \dd_i\left[\dd_y(2\Phi-\Psi) + {1\over 2}\dd_t{\cal B}\right] +
{1\over 2}(\dd_t\dd_y{\Si_i} + \dd_t^2{\cal E}_i - \lap {\cal E}_i), \\
G_{44} & = &\lap (\Psi-2\Phi) +3\dd_t^2\Phi ~.
\end{eqnarray}

The source free  Einstein equations (\ref{sourcefree1}) to
(\ref{sourcefreelast}) are now simply obtained by setting
$G_{AB}=0$. Integrating the five dimensional Einstein equation,
$G_{AB}= 2M_5^{-3}T_{AB}$ over the extra 
dimension $y$, with the ansatz given in Eqs.~(\ref{T4}) to (\ref{Pi}) for the 
stress energy tensor one obtains the 4--dimensional Einstein
equations given in Eqs.~(\ref{eq:scal1'}) to (\ref{eq:tens'}).

}
\end{document}